\documentclass[1p,times]{elsarticle}
\usepackage{graphicx}
\usepackage{lineno}
\graphicspath{{figures/}}
\usepackage{float}
\usepackage{booktabs}
\usepackage{multirow}
\usepackage{amsmath}
\usepackage{hyperref}

\begin{document}

\begin{frontmatter}

\title{A Brief History of Timing}

\author[address1]{N. Cartiglia\corref{mycorrespondingauthor}}
\cortext[mycorrespondingauthor]{Corresponding author}
\ead{cartiglia@to.infn.it}

\address[address1]{INFN sezione di Torino, Torino, Italy}

\begin{abstract}
This review traces the evolution of precision timing in particle physics experiments,
from the first large-scale applications of scintillator and photomultiplier tube (PMT)
systems in the 1990s to the picosecond-precision detectors of future colliders.
Four technological generations are identified: (i) the era of discrete
electronics (NIM, CAMAC, VME) and PMTs, which established the three canonical uses of
timing -- particle identification via time-of-flight, background and pile-up
rejection, and directionality triggering;  (ii) the silicon revolution enabled by Silicon
Photomultipliers (SiPMs), Low-Gain Avalanche Diodes (LGADs), and Application-Specific
Integrated Circuits (ASICs); (iii) the current transition to ubiquitous four-dimensional
(4D) tracking, in which time is a coordinate measured at every point along a particle
trajectory. Under-construction systems at the HL-LHC (CMS MTD, ATLAS HGTD, CMS HGCAL)
demonstrate the maturity of 30--50\,ps silicon timing at the million-channel scale.
The EIC, LHCb VELO Upgrade~II, and ALICE~3 push this technology into new
regimes of radiation hardness, material budget, and granularity.
(iv) The Far-future facilities such as the Muon Collider and FCC require a further leap to
$\sim$10\,ps, setting the agenda for the next decade of detector R\&D.

\end{abstract}


\end{frontmatter}

\tableofcontents

\section{Introduction}
\label{sec:intro}

Timing has always been a tool in high-energy physics experiments, but its role
has changed qualitatively over the past three decades.
In the 1990s, a dedicated timing system measured the transit time of a particle between
two scintillating planes: a single number per track, used offline for particle
identification or event selection.
Today, experiments under construction read out millions of silicon pixels, each
providing both a position and a time-stamp with a precision of 25--50\,ps, and future
facilities aim for a factor of five improvement on this figure.

This transformation was driven by two converging pressures.
On the physics side, the luminosity frontier has produced collision pile-up
densities that can no longer be resolved by spatial tracking alone; at the HL-LHC
an average of 200 proton--proton interactions occur per bunch crossing, yielding to overlapping interactions
along the beam axis but  distributed over a time interval of about  $\sim$150\,ps.
A time-stamp at the track level with a resolution of $\sim 50\, $ps allows the reconstruction algorithms to assign tracks to
primary vertices with high fidelity.
On the technology side, three inventions removed the obstacles that had prevented the
integration of fast timing at the pixel level: the Silicon Photomultiplier (SiPM),
the Low-Gain Avalanche Diode (LGAD), and ASIC able to measure timing with high resolution (timing ASICs). 

The structure of this review follows the historical arc of these developments.
Section~\ref{sec:pmt} covers the era of PMT-based systems and their three canonical
physics applications.
Section~\ref{sec:silicon_revolution} describes the silicon revolution and its enabling
technologies.
Section~\ref{sec:4D} discusses the conceptual shift from a dedicated timing layer to
full 4D tracking.
Sections~\ref{sec:construction}--\ref{sec:future} survey the landscape of systems under
construction and planned for future facilities, organised by experiment and era.
Section~\ref{sec:asics} reviews the corresponding ASIC development.
Sections~\ref{sec:calo}--\ref{sec:gas} cover timing in calorimetry and gaseous/Cherenkov
detectors.
Section~\ref{sec:space} addresses the particular constraints of space-borne experiments.
Section~\ref{sec:challenges} outlines open challenges, and Section~\ref{sec:conclusions}
concludes.

\section{Timing Technologies 1990s--2010s}
\label{sec:pmt}

\subsection{Physics motivations}
\label{sec:pmt_physics}

Before the advent of silicon timing detectors, nearly all timing applications in
particle physics relied on three components: a scintillator pad or bar, a photomultiplier
tube (PMT), and a chain of discrete electronics in NIM, CAMAC, or VME crates.
A Minimum Ionising Particle (MIP) traversing the scintillator deposits energy that is
converted to a burst of optical photons; these are converted to an electrical pulse by
the PMT, which is then digitised and time-stamped by the downstream electronics.
Systems of this type typically achieved time resolutions of 100--200\,ps, occupied
volumes of 1--25\,m$^2$, and required a few hundred PMTs operating at 1--2\,kV.

\begin{figure}[!ht]
  \vspace{.2cm}
  \centering
  \includegraphics[width=.95\linewidth]{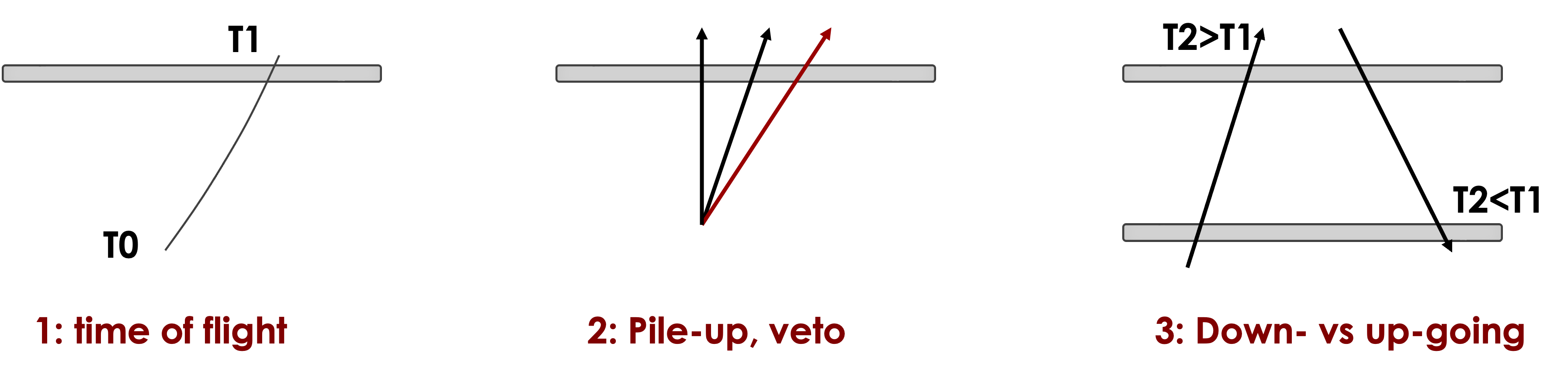}
  \caption{The three canonical uses of timing in particle physics.
  (1)~\textit{Time-of-flight}: a particle traverses a known distance between
  a start counter (T0) and a stop counter (T1); the transit time determines
  its velocity and, combined with momentum, its mass.
  (2)~\textit{Pile-up and veto}: multiple particles arrive at the same detector
  at different times; the out-of-time track (red) is rejected by a timing window.
  (3)~\textit{Directionality}: two stacked detectors measure $T_1$ and $T_2$;
  a downward-going particle satisfies $T_2 > T_1$, while an upward-going particle
  satisfies $T_2 < T_1$.
  \label{fig:three_uses}}
\end{figure}

These systems were deployed for three distinct physics goals, illustrated schematically
in Fig.~\ref{fig:three_uses}:

\paragraph{Particle Identification (PID) via Time-of-Flight (TOF)}
By measuring the transit time of a particle between two detectors separated by a
known distance $L$, one can determine its velocity $\beta = L/(ct)$.
Combined with the momentum $p$ measured by a magnetic spectrometer, the mass
$m^2 = p^2(1/\beta^2-1)$ discriminates between pion, kaon, and proton hypotheses.
For typical momenta of 1--3\,GeV/$c$ and flight paths of 1--2\,m, a resolution of
100\,ps provides $3\sigma$ $\pi/K$ separation up to $\sim$1.5\,GeV/$c$.

\paragraph{Pile-up and background rejection}
In fixed-target and collider experiments with high beam rates, timing resolves
overlapping events in time and suppresses accidental coincidences.
A sub-nanosecond window around the beam bunch eliminates out-of-time activity from
halo particles, photons converted upstream, or interactions of secondary beams.

\paragraph{Directionality and triggering}
The relative arrival times of signals in different detector layers determine whether a
particle is travelling upward or downward, an essential tool for separating cosmic-ray
albedo from genuine upward-going neutrino-induced muons in underground and
space-borne detectors.

\subsection{PMTs - Scintillators systems}

\begin{figure}[!ht]
  \vspace{.2cm}
  \centering
  \includegraphics[width=.48\linewidth]{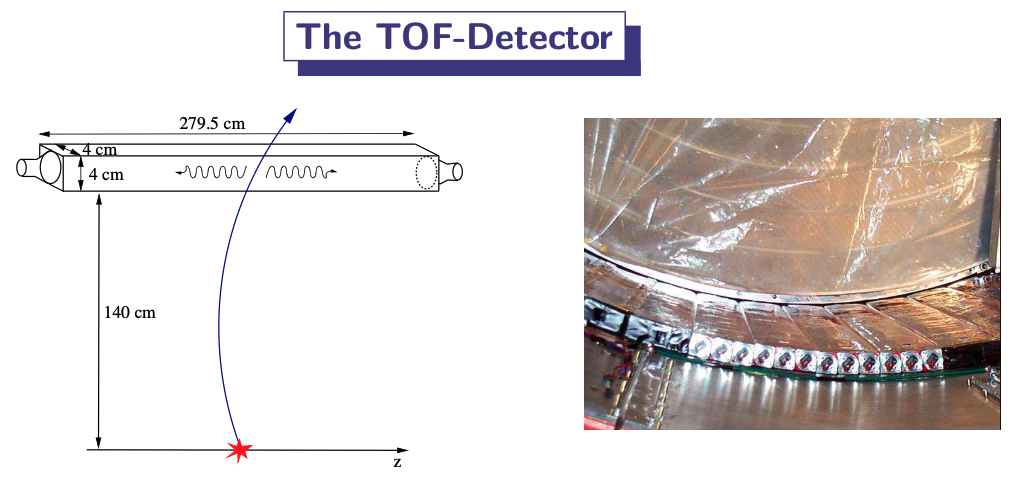}\hfill
  \includegraphics[width=.48\linewidth]{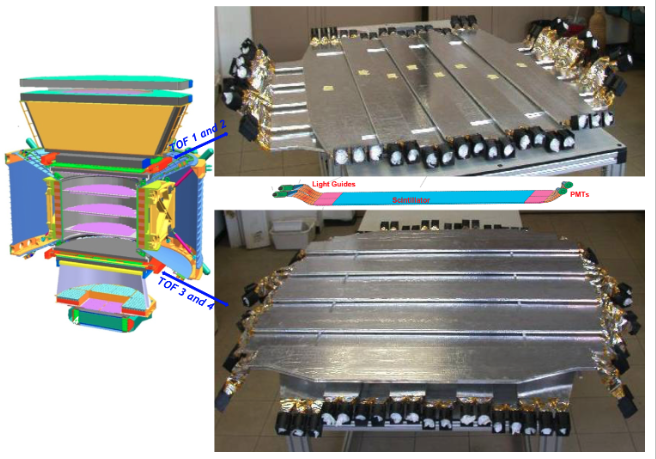}
  \caption{Left: The CDF-II Time-of-Flight detector, showing the cylindrical
  arrangement of scintillator bars (length 279.5\,cm) and fine-mesh PMTs at both
  ends, positioned between the outer tracking volume and the solenoid at a radius
  of 140\,cm~\cite{Acosta2005}.
  Right: One of the four AMS-02 TOF planes, illustrating the scintillator paddles
  with light guides and PMTs.
  \label{fig:pmt_systems}}
\end{figure}

\paragraph{CDF-II Time-of-Flight}
The CDF-II detector at the Tevatron added a TOF barrel system between the central
outer tracker and the solenoid (Fig.~\ref{fig:pmt_systems}, left)~\cite{Acosta2005}.
It comprised 216 scintillator bars of dimension $3\times3\times279$\,cm$^3$,
read out by fine-mesh PMTs capable of operating in the 1.4\,T solenoidal field.
The system achieved a time resolution of $\sim$100\,ps and provided $1.5\,\sigma$
$K/\pi$ separation at 1.6\,GeV/$c$, improving the charged-kaon tagging efficiency
for $B$-meson flavour identification by approximately 20\%.

\paragraph{NA48 and KTeV}
The NA48 experiment at CERN measured direct CP violation in $K_L\to\pi\pi$ decays
by comparing simultaneous $K_L$ and $K_S$ beams~\cite{Fanti1999}.
The Liquid Krypton (LKr) electromagnetic calorimeter provided intrinsic timing at
the $\sim$300\,ps level, supplemented by a plastic-scintillator hodoscope with
$\sim$150\,ps resolution.
Timing was essential to reject accidental associations between the two beams and to
define the fiducial decay volume.
KTeV at Fermilab employed a similar strategy with a CsI calorimeter~\cite{Alavi-Harati1999}.

\paragraph{AMS-02}
The Alpha Magnetic Spectrometer on the International Space Station uses four layers
of plastic-scintillator TOF paddles, read out by PMTs (Fig.~\ref{fig:pmt_systems},
right)~\cite{Aguilar2002}.
The system achieves $\sim$160\,ps time resolution and serves three roles simultaneously:
primary trigger generation for the magnetic spectrometer, charge measurement via $dE/dx$,
and up/down discrimination to separate cosmic-ray albedo protons from genuine upward-going
particles.
Notably, AMS-02 already used custom ASICs in parts of its readout chain, an early
example of the integration trend that would later define silicon timing systems.

\subsection{Water Cherenkov detectors: the Kamiokande legacy}

\begin{figure}[!ht]
  \vspace{.2cm}
  \centering
  \includegraphics[width=.5\linewidth]{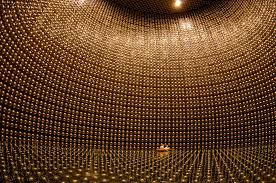}
  \caption{Interior of the Super-Kamiokande detector, showing the array of
  50\,cm diameter PMTs covering the cylindrical walls, floor, and ceiling
  of the 50\,kt water volume. Vertex and direction reconstruction rely entirely
  on the relative arrival times of Cherenkov photons across the PMT array,
  making Super-Kamiokande one of the earliest examples of timing as the primary
  reconstruction tool rather than an auxiliary measurement~\cite{Fukuda1998}.
  \label{fig:superK}}
\end{figure}

Water Cherenkov detectors represent a distinct class in which timing is not an
auxiliary measurement but the \textit{primary reconstruction tool}.
In experiments such as Kamiokande and its successors Super-Kamiokande and
Hyper-Kamiokande~\cite{Fukuda1998} (Fig.~\ref{fig:superK}), the position of a neutrino interaction vertex
and the direction of the resulting charged lepton are reconstructed entirely from the
relative arrival times of Cherenkov photons at the surrounding wall of PMTs.
A photon travelling at speed $c$ from a vertex at position $\vec{r}_0$ arrives at
PMT $i$ at time $t_i = |\vec{r}_i - \vec{r}_0|/c + t_0$.
A fit over all PMTs with at least one detected photon determines $\vec{r}_0$ and $t_0$
with a spatial precision of $\sim$30\,cm for sub-GeV events.
The precision of this reconstruction depends directly on the single-photon transit-time
spread of the PMTs, which in Super-Kamiokande is $\sim$2\,ns for 50-cm diameter tubes,
and on the coverage fraction.

\subsection{The MRPC family: gaseous timing}
\label{sec:mrpc}

\begin{figure}[!ht]
  \vspace{.2cm}
  \centering
  \includegraphics[width=.5\linewidth]{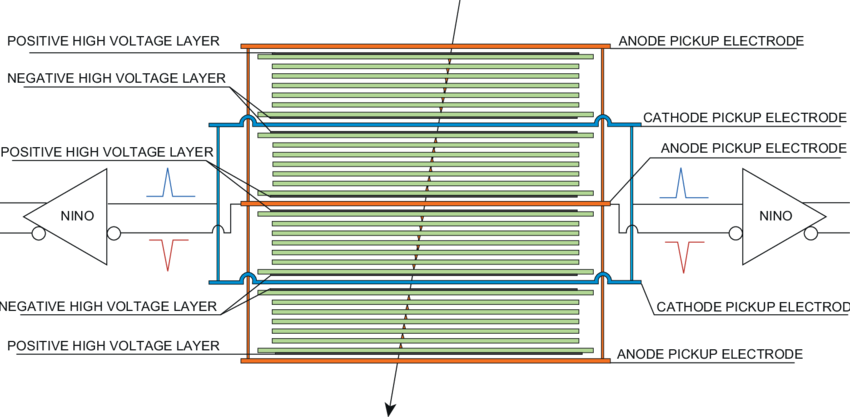}
  \caption{Schematic cross-section of a Multigap Resistive Plate Chamber (MRPC).
  Multiple high-resistivity glass plates (green) are interleaved with sub-millimetre
  gas gaps. The uniform electric field across all gaps simultaneously amplifies
  the avalanche initiated by a traversing particle.
  Differential signals from the anode and cathode pickup electrodes are fed directly
  into the \textsc{NINO} ASIC discriminators (shown at left and right), followed
  by an HPTDC for time digitisation. This readout chain enabled the ALICE TOF to
  achieve $\sim$80\,ps resolution across 140\,m$^2$ of active area~\cite{ALICE_TOF}.
  \label{fig:mrpc}}
\end{figure}

The Multigap Resistive Plate Chamber (MRPC) was developed in the 1990s and enabled a
dramatic increase in the channel density achievable with gaseous timing detectors.
An MRPC consists of a stack of high-resistivity glass plates separated by
sub-millimetre gas gaps.
An applied high voltage ($\sim$15\,kV) creates a uniform electric field across all gaps
simultaneously; a traversing charged particle triggers an avalanche that propagates
across several gaps, producing a fast differential signal.
The key readout electronics were the \textsc{NINO} ASIC~\cite{Anghinolfi2004}, an
ultra-fast differential amplifier-discriminator fabricated in 0.25\,$\mu$m CMOS,
and the HPTDC (High-Performance Time-to-Digital Converter)~\cite{Christiansen2004},
which achieved 25\,ps binning over a large dynamic range (Fig.~\ref{fig:mrpc}).
Together they constituted the first fully ASIC-based timing readout in a major
collider experiment.

The ALICE Time-of-Flight system at the LHC~\cite{ALICE_TOF} deployed this technology
at a scale previously impossible with PMT-based systems:
140\,m$^2$ of active area, 157\,888 readout channels, and a time resolution of $\sim$80\,ps
at the system level (better than 60\,ps for the MRPC strips alone).
The stringent requirements -- large area, high segmentation, low material budget,
and a cost ceiling -- could not have been met by scintillator-PMT technology,
making the ALICE TOF a decisive proof-of-concept for large-scale ASIC-driven timing.

\subsection{The Cherenkov family: DIRC and TOP}
\label{sec:cherenkov_early}

The BaBar~\cite{Aubert2002} and Belle~II~\cite{BelleII_TDR} experiments at the
$B$-factory colliders required compact, low-mass PID systems capable of $\pi/K$
separation in an environment where a conventional TOF approach would have been
physically impossible.
Both solutions exploit the propagation of Cherenkov photons inside a quartz radiator.

The Detection of Internally Reflected Cherenkov light (DIRC) detector of BaBar used
144 fused-silica bars of length 4.9\,m; photons were transported by total internal
reflection to a water-filled standoff box instrumented with conventional PMTs,
achieving $\sim$2\,mrad angular resolution per photon.

The Time-of-Propagation (TOP) counter of Belle~II measures both the arrival time and
the lateral position of Cherenkov photons at the end of a 2.5\,m quartz plate read
out by micro-channel plate PMTs with $\sim$100\,ps single-photon resolution.
Since the path length inside the quartz depends on the Cherenkov angle and hence on
the particle velocity, a single time measurement encodes the Cherenkov angle.
Pions and kaons of the same momentum travel at slightly different velocities,
producing photons at slightly different angles and therefore arriving with different
delays; with a per-track time resolution below 50\,ps the TOP counter provides
$>3\sigma$ $K/\pi$ separation up to $\sim$4\,GeV/$c$.

\label{sec:pmt_summary}

Table~\ref{tab:pmt_summary} summarises the principal timing systems of this era.
The common features are: $\mathcal{O}(100)$\,ps resolution, $\mathcal{O}(100)$--$\mathcal{O}(10^5)$
readout channels, and a reliance on PMTs as the photosensor.
The MRPC/NINO/HPTDC line pioneered ASIC-based readout and large-scale gaseous
timing, presaging the silicon era.

\begin{table}[t!]
\vspace{.2cm}
\begin{center}
\begin{tabular}{l l l l l}
\toprule
Experiment & Technology & Area / Channels & Resolution & Primary use \\
\midrule
CDF-II TOF     & Scint.\ + fine-mesh PMT  & 216 bars             & $\sim$100\,ps  & PID ($K/\pi$) \\
NA48 hodoscope & Scint.\ + PMT            & $\sim$150 channels   & $\sim$150\,ps  & Pile-up rejection \\
AMS-02 TOF     & Scint.\ + PMT            & 34 paddles           & $\sim$160\,ps  & Trigger, directionality \\
ALICE TOF      & MRPC (gaseous)           & 140\,m$^2$, 157\,888 & $\sim$80\,ps   & Hadron PID \\
Belle~II TOP   & Quartz + MCP-PMT         & 16 modules           & $<$50\,ps/track & PID ($K/\pi$) \\
Super-K PMT wall & Water Cherenkov + PMT  & 11\,146 PMTs         & $\sim$2\,ns/PMT & Vertex, direction \\
\bottomrule
\end{tabular}
\caption{Representative timing systems of the 1990s and 2000s. The MRPC-based ALICE TOF
was the first large-scale ASIC-driven timing system and bridged the gap to the
silicon era.
\label{tab:pmt_summary}}
\end{center}
\vspace{-.3cm}
\end{table}

\section{The Silicon Revolution}
\label{sec:silicon_revolution}

The limitations of PMT-based systems -- bulk, fragility, high-voltage requirements,
sensitivity to magnetic fields, and limited granularity -- became incompatible with
the requirements of the next generation of experiments.
Three inventions, developed in the 2000s and 2010s, removed
these constraints: the SiPM for photon detection, the LGAD for direct charged-particle
detection, and the deep-submicron timing ASIC.

\subsection{Silicon Photomultiplier (SiPM)}
\label{sec:sipm}

\begin{figure}[!ht]
  \vspace{.2cm}
  \centering
  \includegraphics[width=.5\linewidth]{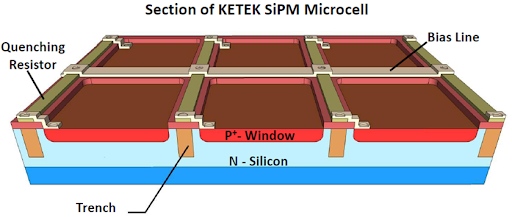}
  \caption{Three-dimensional cross-section of a KETEK SiPM microcell, showing the
  individual SPAD structure: the P$^+$ window (active region) on N-silicon bulk,
  the trench isolation between cells, and the bias/quenching resistor network on
  the surface. Each microcell fires independently in Geiger mode; the parallel
  sum of all cells gives an analog output proportional to the number of incident
  photons. Modern SiPMs integrate $10^3$--$10^4$ such cells per mm$^2$.
  \label{fig:sipm}}
\end{figure}

A Silicon Photomultiplier is an array of tens to hundreds of thousands of Single-Photon
Avalanche Diode (SPAD) microcells, each operating in Geiger mode, connected in parallel
on a common substrate~\cite{Renker2009} (Fig.~\ref{fig:sipm}).
When a photon liberates an electron-hole pair in a SPAD biased a few volts above
breakdown, the resulting Geiger discharge produces a stereotyped charge pulse of
$\sim$10$^6$ electrons in $\sim$1\,ns.
The sum of all firing cells constitutes an analog output proportional to the number of
incident photons, provided the number of photons is much less than the number of cells.

SiPMs match or exceed the photon-detection efficiency of bialkali PMTs ($>$40\% peak),
while operating at only 30--60\,V bias compared to the 1--2\,kV required by PMTs.
Their millimetre-scale footprint allows tight coupling to crystal scintillators,
they are immune to magnetic fields, and their unit cost at volume has fallen
dramatically with CMOS foundry production.
These properties make them the photosensor of choice for scintillator-based timing
systems at the HL-LHC and beyond.

\paragraph{Scintillation and Cherenkov light with SiPMs}
When a charged particle traverses a dense medium it can produce two distinct
types of light, each with different temporal characteristics and timing implications.
\textit{Scintillation light} is emitted isotropically with a characteristic
fluorescence lifetime of the crystal (a few nanoseconds for LYSO, $\sim$40\,ns
for BGO), and is therefore delayed and spread in time.
\textit{Cherenkov light} is emitted promptly ($<$1\,ps) in a forward cone at
angle $\cos\theta_C = 1/n\beta$, and represents a nearly instantaneous signal
that can, in principle, reach sub-100\,ps timing even from a single detected photon.

\begin{figure}[!ht]
  \vspace{.2cm}
  \centering
  \includegraphics[width=.90\linewidth]{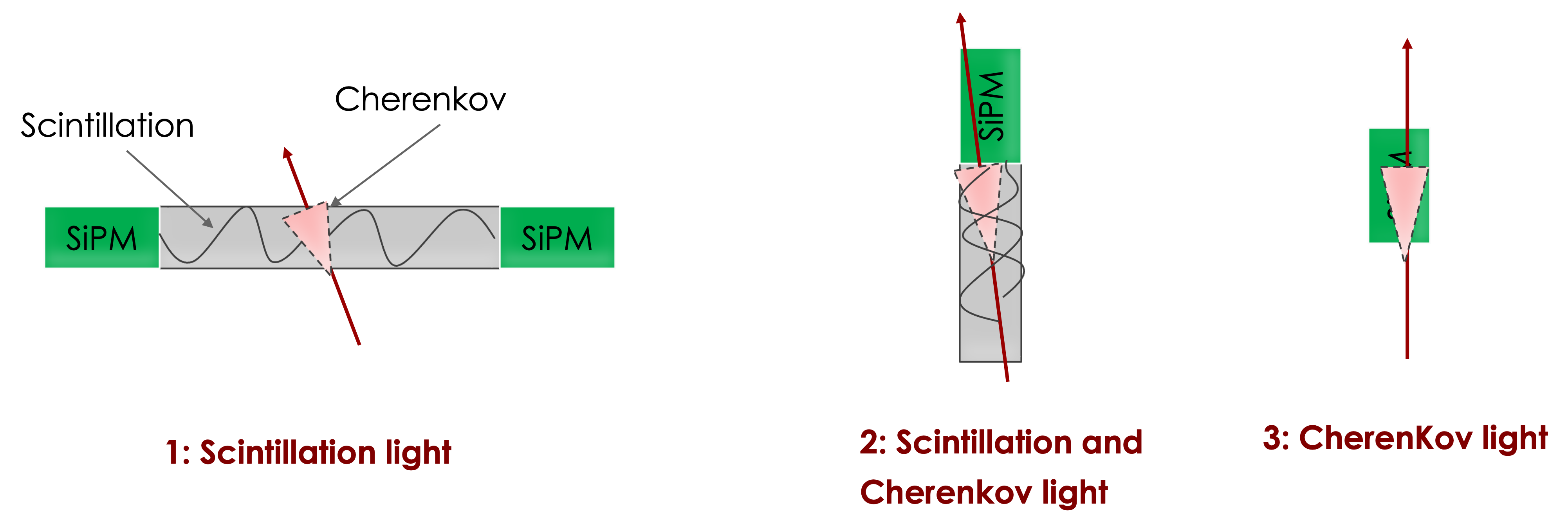}
  \caption{Three configurations for SiPM-based optical timing.
  (1)~\textit{Scintillation only}: SiPMs on the lateral sides of a crystal bar
  collect isotropic scintillation light after multiple reflections; the Cherenkov
  cone directed along the bar axis is not collected.
  This geometry maximises the number of detected photons but the timing is
  limited by the crystal fluorescence lifetime.
  (2)~\textit{Scintillation and Cherenkov}: the SiPM is placed at the end of the
  crystal, along the particle direction; forward Cherenkov photons arrive promptly
  and directly, providing a fast timing signal, while scintillation photons
  arrive later and give the energy measurement.
  (3)~\textit{Cherenkov only}: Cherenkov light produced directly in the SiPM
  package or a thin radiator provides a purely prompt signal with very few
  photons, representing the ultimate timing limit of the technology.
  \label{fig:sipm_light}}
\end{figure}

Three readout geometries exploit these two components differently,
as illustrated in Fig.~\ref{fig:sipm_light}:

\begin{enumerate}
  \item Scintillation only.
  SiPMs are mounted on the lateral faces of a crystal bar.
  Scintillation light bounces along the bar and reaches the SiPMs after multiple
  reflections, producing a large but temporally broadened signal.
  The Cherenkov cone is directed along the particle trajectory toward the bar ends
  and is not collected; the timing is dominated by the crystal decay time.
  
  \item Scintillation and Cherenkov.
  The SiPM is placed at the end of the crystal, aligned with the particle
  direction.
  Cherenkov photons emitted in the forward cone arrive at the SiPM promptly and
  directly; scintillation photons arrive later after diffuse propagation.
  The Cherenkov component provides a fast leading edge, while the slower
  scintillation component yields the energy measurement.
  This dual-component signal is the basis of precision timing in
  PbWO$_4$ and PbF$_2$ crystals.

  \item Cherenkov only in the SiPM.
  At sufficiently high particle velocity, Cherenkov light is generated directly
  inside the SiPM silicon substrate or its protective window, without any external
  crystal.
  Only a handful of photons are produced, but they arrive with sub-picosecond
  jitter.
  This effect, while difficult to exploit at low energies, sets an intrinsic
  physical lower bound on the timing resolution achievable with SiPM-based
  detectors.
\end{enumerate}

The principal limitations of SiPMs for timing are correlated noise (after-pulsing,
optical cross-talk) and the capacitance of large-area devices, which degrades the
signal-to-noise ratio when the cell count is large.
The latest SiPM productions have a
reduced dark-count rates, $\sim$100\,kcps/mm$^2$ at 20$^\circ$C, an important
figure for the low-photon-count regime of Cherenkov readout.

\subsection{Low-Gain Avalanche Diode (LGAD / UFSD)}
\label{sec:lgad}

\begin{figure}[!ht]
  \vspace{.2cm}
  \centering
  \includegraphics[width=.6\linewidth]{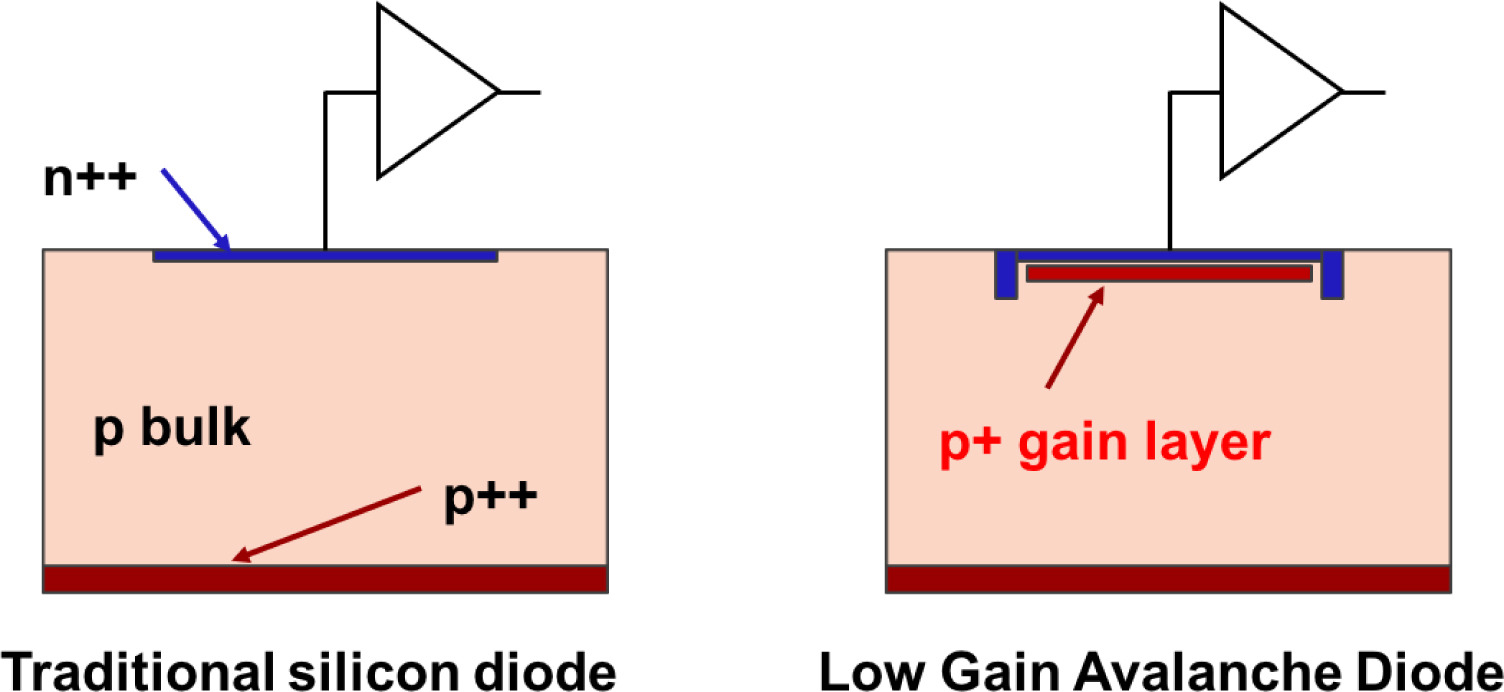}
  \caption{Schematic cross-sections comparing a traditional silicon diode (left)
  and a Low-Gain Avalanche Diode (right). In the LGAD, a thin p$^+$ gain layer
  implanted just below the n$^{++}$ electrode creates a localised high-field
  region that provides internal multiplication of factor 10--30. This moderate
  gain is sufficient to produce fast current pulses with excellent
  signal-to-noise ratio, enabling time resolutions of 25--40\,ps per
  layer~\cite{Pellegrini2014}.
  \label{fig:lgad_xsec}}
\end{figure}

While SiPMs detect photons via Geiger-mode operation, a different silicon structure is
required to detect the direct ionisation signal from a charged particle.
The Low-Gain Avalanche Diode (LGAD) design \cite{Pellegrini2014} (Fig.~\ref{fig:lgad_xsec}) is the key technological step that allows the design of silicon sensors with high temporal resolution, 
also known as Ultra-Fast Silicon Detector (UFSD)~\cite{Cartiglia2015}. This technology uses a thin implanted $p^+$ gain layer to provide an internal
multiplication factor of 10--30.

A MIP traversing a 50--300\,$\mu$m thick LGAD deposits $\sim$70--2000\,electron-hole
pairs. Internal multiplication raises this to a collected charge of 5--20\,fC, producing a
fast current pulse with a rise time below 1\,ns.
The combination of high signal amplitude and fast rise time yields an excellent
signal-to-noise ratio and, critically, allowed to reduce the electronic jitter below 30\,ps~\cite{Sadrozinski2018}.

The temporal resolution $\sigma_t$ of an UFSD is decomposed into two independent contributions~\cite{Sadrozinski2018}:
\begin{equation}
  \sigma_t^2 = \sigma_\text{jitter}^2 + \sigma_\text{Landau Noise}^2
\end{equation}

\begin{itemize}
\item Jitter arises from electronic noise $\sigma_N$ and the signal slew rate $dV/dt$ at threshold crossing, $ \sigma_\text{jitter} = \frac{\sigma_N}{dV/dt}$. It dominates at low gain, where the signal amplitude is small relative to the noise.
\item Landau noise, also called the non-uniform ionization term, arises from the event-to-event variability of the charge deposition. Each MIP deposits charge in a different spatial pattern along the track, producing a different signal shape and hence a different trigger time. This term dominates at high gain, where jitter is negligible. The Landau noise  grows with the square root of the thickness $d$ and decreases with drift velocity $v$,  $ \sigma_\text{Landau Noise} \propto \sqrt{d}/v$. Operating at fields sufficient to saturate the drift velocity $v$, and reducing sensor thickness, are the most effective ways to reduce this term.  Sensors with a thickness is in the range 35--55\,$\mu$m  have an intrinsic resolutions of 15--25\,ps.
\end{itemize}

Radiation hardness is a critical parameter for collider applications.
The gain layer is composed of boron acceptors that are progressively deactivated by
proton and neutron irradiation through the acceptor removal mechanism~\cite{Moll2018}.
UFSDs retain useful gain up to fluences of $\sim 2\times10^{15}$\,n$_\text{eq}$/cm$^2$,
sufficient for the HL-LHC appplications; higher fluence applications require
alternative geometries described below.

\subsection{Resistive Silicon Detector (RSD / AC-LGAD)}
\label{sec:rsd}

\begin{figure}[!ht]
  \vspace{.2cm}
  \centering
  \includegraphics[width=.65\linewidth]{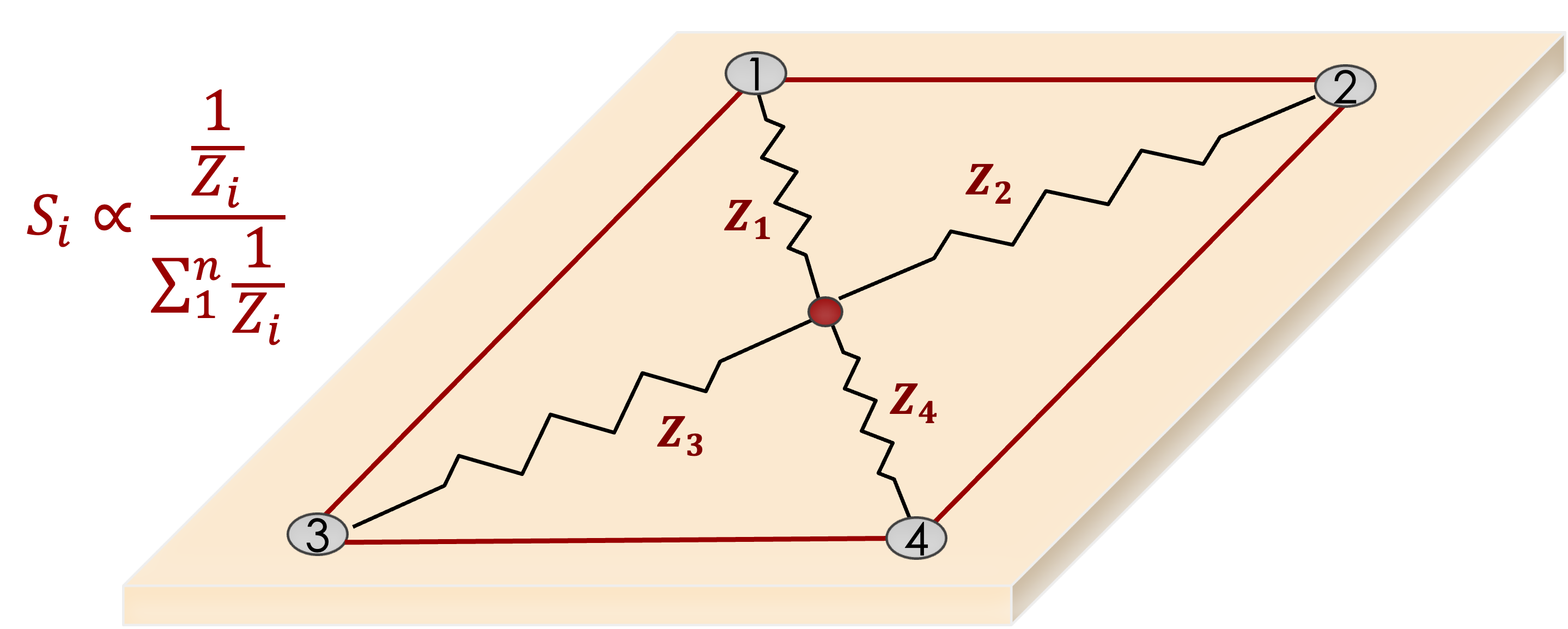}
  \caption{Operating principle of the Resistive Silicon Detector (RSD).
  A MIP hits the continuous resistive surface at the red dot, and the induced
  signal spreads to the four corner electrodes (1--4) through the resistive
  network with impedances $Z_1$--$Z_4$.
  The fraction of the total charge collected by each electrode is
  $S_i \propto (1/Z_i)/\sum_k(1/Z_k)$, where $Z_i$ is proportional to the
  distance from the hit to electrode $i$.
  \label{fig:rsd}}
\end{figure}

The Resistive Silicon Detectors (RSDs)  are an UFSD variant in which the
$n^+$ electrode is replaced by a continuous resistive layer, and the readout pads are connected in AC or DC mode
to the surface~\cite{Mandurrino2019}.
When a MIP hits the sensor, the signal spreads laterally under the resistive layer
before being read by the nearest electrodes. By measuring the charge fraction on each electrode, the hit position can be
reconstructed with a precision below 5\% of the pixel size, far better than the
electrode pitch. The position reconstruction relies on the fact that the resistive layer acts as a
two-dimensional impedance network.
When a MIP hits the sensor at a point surrounded by $n$ readout electrodes,
each at impedance distance $Z_i$ from the hit, the signal fraction collected by
electrode $i$ is
\begin{equation}
  S_i \propto \frac{1/Z_i}{\displaystyle\sum_{k=1}^{n} 1/Z_k},
\end{equation}
as illustrated in Fig.~\ref{fig:rsd}.

RSD thus decouples the spatial resolution from the electrode granularity, using a fraction of the power and readout
complexity of an equivalent array of small standard LGADs. This property makes RSD attractive for applications where the power budget or the
number of readout channels must be minimised.

Table~\ref{tab:silicon_summary} present a summary of the silicon sensors used for timing. 

\begin{table}[t!]
\vspace{.2cm}
\begin{center}
\begin{tabular}{l l l l}
\toprule
Technology & Detection & Gain & Timing resolution \\
\midrule
SiPM                & Photon (indirect)   & $10^5$--$10^6$ (Geiger) & $<$100\,ps / photon \\
LGAD / UFSD         & MIP (direct)        & 10--30 (linear)         & 25--40\,ps / layer  \\
RSD / AC-LGAD       & MIP (direct)        & 10--30 (linear)         & 25--40\,ps / layer  \\
Standard silicon    & MIP (direct)        & 1 (no gain)             & $>$200\,ps          \\
3D trench silicon   & MIP (direct)        & 1--low                  & $\sim$10\,ps (lab)  \\
\bottomrule
\end{tabular}
\caption{Silicon sensors presently used for timing.
\label{tab:silicon_summary}}
\end{center}
\vspace{-.3cm}
\end{table}

\section{From 3+1 to 4D Tracking}
\label{sec:4D}

\subsection{The conceptual shift}

In all the experiments discussed so far, timing was a \textit{dedicated} measurement
confined to a specific detector subsystem: a timing layer, a TOF wall, a hodoscope.
The tracking detector itself provided only position information, and timing was added
as a separate coordinate, useful but not integrated into the event reconstruction.

The availability of SiPMs, LGADs, and timing ASICs makes a qualitatively
different use of timing possible.
The concept of 4D tracking adds time as a fourth coordinate measured at
\textit{every point} along the track, alongside the three spatial coordinates.
This is to be distinguished from the simpler 3+1 tracking, in which a timing
layer provides a single time measurement per track, assigned during final reconstruction.

\subsection{The 4D tracking challenge}
\label{sec:4D_challenge}
\begin{figure}[!ht]
  \vspace{.2cm}
  \centering
  \includegraphics[width=.95\linewidth]{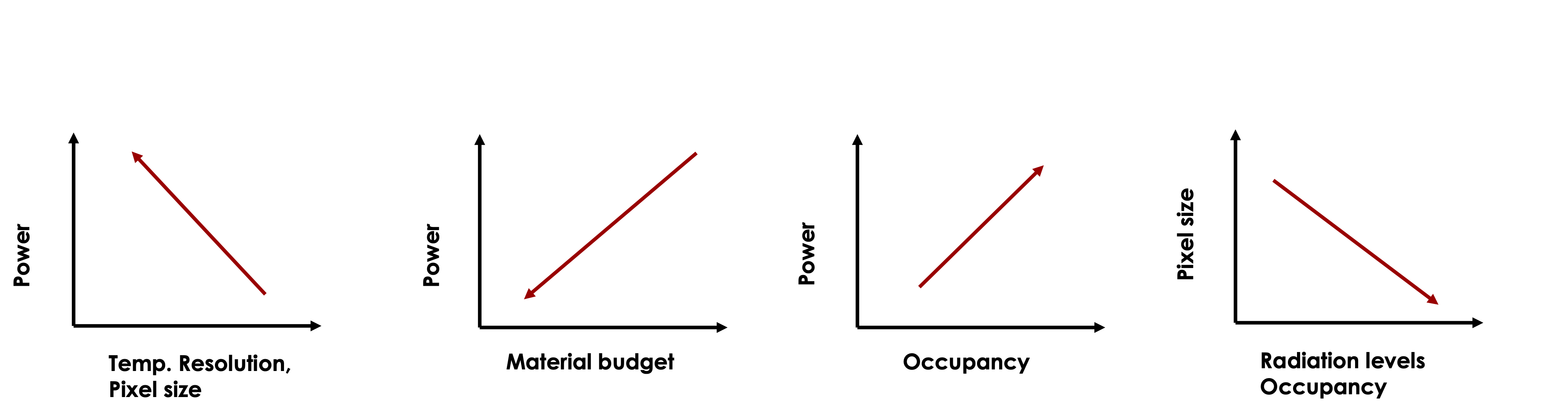}
  \caption{The four fundamental trade-offs in the optimisation of a 4D silicon
  tracking sensor.
  \textit{First}: power consumption increases as timing resolution improves
  and pixel size decreases -- faster front-end amplifiers and higher-frequency
  TDC oscillators consume more current, and finer granularity means more channels
  per unit area.
  \textit{Second}: higher power requires more cooling infrastructure
  (CO$_2$ circuits, support structures), directly increasing the material budget
  and therefore the multiple-scattering contribution to spatial resolution.
  \textit{Third}: higher occupancy environments demand more power, because
  front-end circuits must process hits at higher rates without dead time.
  \textit{Fourth}: harsher radiation environments and higher occupancy force
  the use of smaller pixels to limit per-pixel hit rates and to maintain
  spatial resolution after radiation damage, but smaller pixels amplify the
  power and material problems.
  Each experiment sits at a different point in this multi-dimensional space,
  making a universal 4D solution impossible.
  \label{fig:compromise}}
\end{figure}

The optimisation of a 4D system is not a single-variable problem.
Five parameters must be balanced simultaneously: temporal resolution, spatial
resolution and pixel size, material budget, occupancy, and radiation level.
These parameters are deeply intertwined, and improving one invariably degrades
at least one other.
As a consequence, no single sensor technology is optimal for all applications,
and the landscape of 4D detectors is necessarily diverse.
Figure~\ref{fig:compromise} illustrates the four dominant trade-offs.

\paragraph{Power and timing resolution}
The ASIC power budget is the central bottleneck.
Achieving good timing requires a fast, low-noise amplifier (large bias current)
and a high-frequency TDC (large dynamic power).
In 28\,nm CMOS, a per-pixel TDC operating at the $\sim$10\,ps level consumes
$\sim$1.5--2\,W/cm$^2$.
The maximum power that can be removed by an evaporative CO$_2$ cooling circuit
with acceptable material budget is $\sim$0.3--0.5\,W/cm$^2$, a shortfall of a factor
four to six that is the dominant engineering challenge for next-generation trackers.
Relaxing the timing requirement to $\sim$25--30\,ps (HL-LHC level) reduces the
power to a manageable $\sim$0.2--0.4\,W/cm$^2$, which is why HL-LHC timing layers
are feasible today whereas a full 4D inner tracker is not yet.

\paragraph{Power and material budget}
Every watt of heat dissipated in the detector volume requires a cooling circuit.
A standard evaporative CO$_2$ system contributes $\sim$0.5--1\%\,$X_0$ per
cooling loop, which is comparable to or larger than the sensor itself.
For inner tracker layers, where the total material budget target is often
$\mathcal{O}(1\%)$\,$X_0$ per layer, the cooling infrastructure can dominate
the material budget and undo the gains from using thin sensors.
This coupling between power and material is one of the strongest arguments for
the monolithic CMOS-LGAD approach (Section~\ref{sec:alice3}), where the lower
power density makes air cooling viable and eliminates the CO$_2$ circuit entirely.

\paragraph{Occupancy and power}
High-occupancy environments -- the inner layers of high-luminosity collider
detectors -- require front-end electronics that can sustain high hit rates
without saturation or dead time.
Sustaining low dead time at high hit rates requires larger buffers, faster
serialisers, and higher-bandwidth data links, all of which increase power.
The LHCb VELO Upgrade~II inner region, with $\sim$800 tracks per crossing at
5\,mm from the beam, represents the current extreme of this trade-off.

\paragraph{Radiation, occupancy, and pixel size}
Radiation damage degrades the charge collection efficiency and increases the
leakage current of silicon sensors; both effects worsen as the fluence increases.
Reducing the pixel area limits the leakage current per channel (improving the
signal-to-noise ratio) and reduces the per-pixel occupancy, preserving spatial
resolution.
However, smaller pixels mean more channels per unit area, more ASICs, more power,
and more material.
The minimum viable pixel size is therefore set by a balance between the radiation
hardness requirement and the power/material budget, a balance that shifts with
every new ASIC process node.
At present, $55\times55\,\mu$m$^2$ at 28\,nm represents the practical frontier
for inner vertex detectors at the LHC (LHCb VELO Upgrade~II); future colliders
will likely require $20\times20\,\mu$m$^2$ or below, motivating the move to
14\,nm or 7\,nm CMOS processes.

\section{Timing Systems Under Construction}
\label{sec:construction}

\subsection{CMS MIP Timing Detector (MTD)}
\label{sec:cms_mtd}

The CMS experiment at the HL-LHC will install the MIP Timing Detector (MTD) in
Phase-2~\cite{CMS_MTD_TDR}.
The MTD provides one time measurement per track (3+1 mode) for pile-up rejection
at $\langle\mu\rangle = 200$.
The detector occupies 4.5\,cm of radial space in the barrel (Barrel Timing Layer, BTL)
and 9.9\,cm of longitudinal space in the endcap (Endcap Timing Layer, ETL).

\paragraph{Barrel Timing Layer (BTL)}
The BTL uses LYSO:Ce crystal scintillator bars ($3.75\times3.75\times57$\,mm$^3$)
read out by SiPMs on both ends.
The total active area is $\sim$38\,m$^2$ with $\sim$350\,000 channels.
The TOFHIR2 ASIC~\cite{TOFHIR2} (130\,nm CMOS) provides ToA and charge measurements
and targets a time resolution of 30\,ps (beginning of life) to 60\,ps (end of life,
after irradiation degrades the crystal light yield).

\paragraph{Endcap Timing Layer (ETL)}
The ETL uses LGAD sensors of $1.3\times1.3$\,mm$^2$ pixel size in a double-disk
geometry, covering $\sim$14\,m$^2$ with $\sim$8.5\,million channels.
The ETROC ASIC~\cite{ETROC} (65\,nm CMOS) provides ToA with ToT-based time-walk
correction and targets $\sim$25\,ps electronics jitter, contributing to a
system-level goal of $\sim$30--40\,ps per track.
Evaporated CO$_2$ cooling manages the ASIC power density within the tight material
budget allowed in the tracker volume.

\subsection{ATLAS High-Granularity Timing Detector (HGTD)}
\label{sec:atlas_hgtd}

The ATLAS Phase-2 upgrade includes the High-Granularity Timing Detector
(HGTD)~\cite{ATLAS_HGTD_TDR}, positioned in front of the endcap calorimeter at
$|z|=3.5$\,m.
It covers $2.4<|\eta|<4.0$ with LGAD pixels of $1.3\times1.3$\,mm$^2$, for a total active
area of $\sim$6.4\,m$^2$ and $\sim$3.5\,million channels.
The ALTIROC ASIC~\cite{ALTIROC} (130\,nm CMOS) provides a target time resolution of $\sim$35\,ps per
layer, read out with two sensor layers per disk for a total per-track resolution of
$\sim$25\,ps after combination.
Like the CMS ETL, the HGTD uses evaporated CO$_2$ cooling integrated within
the detector structure.

\subsection{CMS High-Granularity Calorimeter (HGCAL)}
\label{sec:hgcal_overview}

\begin{figure}[!ht]
  \vspace{.2cm}
  \centering
  \includegraphics[width=.90\linewidth]{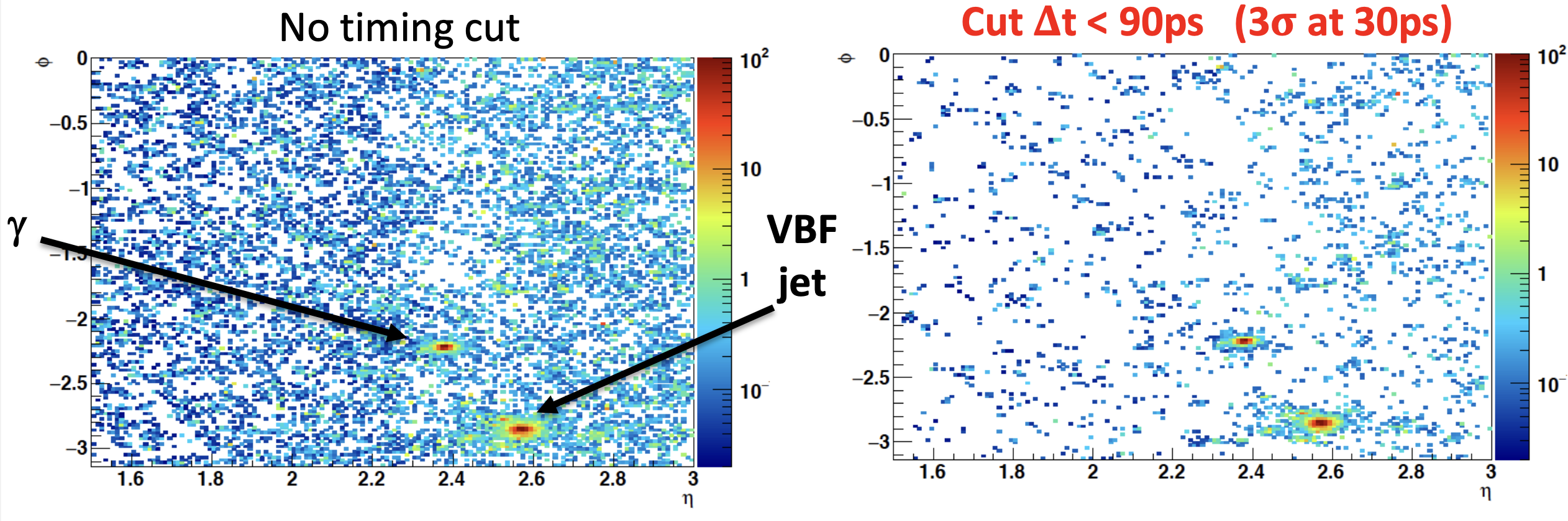}
  \caption{Effect of a 90\,ps timing cut ($3\sigma$ at 30\,ps resolution) on
  HGCAL occupancy. Left: the $(\eta, \phi)$ map of energy deposits without any
  timing selection, showing a dense background of out-of-time pile-up hits
  throughout the endcap. Right: after applying $|\Delta t| < 90$\,ps, the
  pile-up halo is almost entirely removed while the in-time VBF jet and
  photon candidates are preserved~\cite{CMS_HGCAL_TDR}.
  \label{fig:hgcal_timing}}
\end{figure}

The CMS HGCAL~\cite{CMS_HGCAL_TDR} replaces the existing endcap calorimeter with a
sampling imaging calorimeter consisting of 47 layers alternating silicon sensors
and scintillator tiles read out by SiPMs.
In the electromagnetic section, the active elements are hexagonal silicon sensors of
$0.5$--$1.1$\,cm$^2$ cell size; in the hadronic section, SiPM-on-scintillator tiles
provide timing alongside energy measurement.
The timing performance goal is $\sim$30\,ps for electromagnetic clusters with energy
above 5\,GeV, sufficient to reject out-of-time pile-up and to associate electromagnetic
showers with the correct primary vertex.
The primary function is rejecting ``halo'' hits -- energy deposits from out-of-time
pile-up particles that arrive at the calorimeter with a time offset of $\pm$3\,ns.
A 30\,ps timing cut retains $>$99\% of genuine in-time showers while suppressing the
pile-up contribution, improving jet energy resolution in the 200-pile-up environment
by $\sim$20\%, as shown in Fig.~\ref{fig:hgcal_timing}.

The key parameters of all HL-LHC timing systems are collected in Table~\ref{tab:construction}.

\begin{table}[t!]
\vspace{.2cm}
\begin{center}
\begin{tabular}{l l l l l l}
\toprule
System & Technology & Area & Channels & $\sigma_t$ goal  \\
\midrule
CMS MTD BTL  & LYSO + SiPM + TOFHIR2   & 38\,m$^2$  & 350\,k   & 30--60\,ps  \\
CMS MTD ETL  & LGAD + ETROC (65\,nm)   & 14\,m$^2$  & 8.5\,M   & 30--40\,ps  \\
ATLAS HGTD   & LGAD + ALTIROC (130\,nm)& 6.4\,m$^2$ & 3.5\,M   & 25--35\,ps  \\
CMS HGCAL    & Si + SiPM tiles + HGCROC         & endcap     & $\sim$6\,M & $\sim$30\,ps (cluster)  \\
\bottomrule
\end{tabular}
\caption{Timing systems under construction for the HL-LHC (installation $\sim$2028--2030).
\label{tab:construction}}
\end{center}
\vspace{-.3cm}
\end{table}

\section{Timing in New and Near-Future Experiments}
\label{sec:new_experiments}

\subsection{PIONEER at PSI}
\label{sec:pioneer}

PIONEER~\cite{PIONEER_CDR} is a proposed rare pion decay experiment at the Paul
Scherrer Institute, aiming to measure $\mathcal{B}(\pi^+\to e^+\nu)/ \mathcal{B}(\pi^+\to\mu^+\nu)$
at 0.01\% precision, a factor of fifteen improvement on the current world average.
The central challenge is suppressing the $\pi\to\mu\to e$ decay chain background,
in which the pion stops and decays to a muon, which in turn decays to a positron, all
within a few hundred picoseconds and within the same spatial volume of the active target.

The Active TARget (ATAR) of PIONEER uses an AC-LGAD (RSD) stack to perform
decay-chain topology reconstruction in four dimensions.
With a per-layer time resolution of $<$100\,ps, the system distinguishes the stopping
$\pi^+$ signal from the subsequent $\mu^+$ signal and the final $e^+$ signal as
separate hits displaced in time even when they are spatially coincident.
The readout ASIC is FAST3~\cite{FAST3}, a front-end with $\sim$30\,ps resolution designed to read out thin LGADs, providing ToA and charge measurements.

\subsection{Electron-Ion Collider (EIC / ePIC)}
\label{sec:eic}

The Electron-Ion Collider at Brookhaven National Laboratory~\cite{EIC_CDR} will begin
operation in the early 2030s, using the ePIC detector for its first physics programme.
Timing is essential for particle identification across a wide momentum range in the
hadronic final state.

The ePIC Time-of-Flight system uses AC-LGAD (RSD) sensors for the barrel and forward
regions.
The barrel TOF consists of RSD strips targeting a time resolution of 25--35\,ps, read
out by the EICROC ASIC (130\,nm CMOS)~\cite{EICROC}.
The forward TOF uses $0.5\times0.5$\,mm$^2$ RSD pixels for high-rate 4D tracking.
Together these provide $3\sigma$ $\pi/K/p$ separation over the relevant momentum range,
with a flight path of $\sim$1\,m.
EICROC measures both Time-of-Arrival (ToA) and pulse amplitude (charge), the latter
being essential for the position reconstruction algorithm of AC-LGAD sensors.

\subsection{LHCb VELO Upgrade II: the first 4D vertex detector}
\label{sec:lhcb}

The LHCb upgrade for Long Shutdown 4 ($\sim$2034) will replace the current Vertex
Locator with a full 4D tracker -- the first application of per-hit timing in a
collider vertex detector~\cite{LHCb_VELO_II}.
The operating environment is extreme:
radiation fluence up to $6\times10^{16}$\,n$_\text{eq}$/cm$^2$ at 5\,mm from the beam,
very high track occupancy ($\sim$800 tracks per crossing in the inner region), and a
strict power budget.
The total active area is $\sim$0.15\,m$^2$.

The sensor technology uses 3D Trench Silicon sensors for the innermost region,
where LGAD gain would be entirely lost by radiation damage, and LGAD or planar sensors
for the outer region.
Pixel size is $55\times55\,\mu$m$^2$.
The PICOPIX ASIC (28\,nm CMOS)~\cite{PICOPIX} implements the \textit{Analogue Island}
concept, in which analogue shaping and TDC resources are shared among clusters of
pixels, reducing the per-pixel power below the threshold sustainable by the CO$_2$
evaporative cooling, while maintaining a time resolution of $\sim$20\,ps.
With 7--10 timing layers per track, the VELO Upgrade~II will perform full 4D vertexing
with a vertex time resolution of $\sim$7--10\,ps.

\subsection{ALICE 3}
\label{sec:alice3}

ALICE~3~\cite{ALICE3_LoI} represents the opposite extreme from LHCb: a nearly massless
4D tracker for a low-radiation, large-area environment.
The physics programme requires pion-kaon separation at momenta below 1\,GeV/$c$,
demanding TOF measurements with a system resolution approaching 20\,ps over a
flight path of a few centimetres.
The total active area is $\sim$45\,m$^2$.

The sensor concept is a Monolithic CMOS-LGAD (for example, the ARCADIA or MADPix projects), in
which the gain layer and readout electronics are integrated on the same silicon die
using a commercial CMOS process.
The monolithic approach avoids the bump-bonding step needed for hybrid assemblies,
reducing the material budget to $\mathcal{O}(0.1\%)\,X_0$ per layer.
Power consumption is targeted below 40\,mW/cm$^2$, feasible with air cooling,
enabling the cylindrical bent geometry required by the ALICE~3 design.
Table~\ref{tab:new_experiments} summarises the key parameters of these detectors.

\begin{table}[t!]
\vspace{.2cm}
\begin{center}
\begin{tabular}{l l l l l}
\toprule
Experiment & Technology & Mode & $\sigma_t$ goal & Key challenge \\
\midrule
PIONEER (PSI)     & AC-LGAD (RSD)       & 4D active target & $<$100\,ps       & Decay chain separation \\
ePIC (EIC)        & AC-LGAD (RSD)       & TOF + 4D         & 25--35\,ps       & PID, momentum range \\
LHCb VELO Upgrade II & 3D Trench + LGAD  & 4D vertex        & $\sim$20\,ps/hit & Radiation, power \\
ALICE 3           & Monolithic CMOS-LGAD& 4D tracking      & $\sim$20\,ps     & Material, area, power \\
\bottomrule
\end{tabular}
\caption{Silicon timing detectors for new and near-future experiments.
\label{tab:new_experiments}}
\end{center}
\vspace{-.3cm}
\end{table}

\section{Timing in Far-Future Experiments}
\label{sec:future}

The far-future experiments at proposed facilities push timing requirements to their
physical limits, demanding a further factor of three to five improvement in per-hit
resolution compared to current HL-LHC systems.

\subsection{FCC-ee}
\label{sec:fcce}

The Future Circular Collider in the electron-positron mode (FCC-ee)~\cite{FCCee_CDR}
is a precision electroweak, flavour, and Higgs factory.
The TOF system targets $\sim$10\,ps per track for $3\sigma$ $K/\pi$ separation up to
$\sim$5\,GeV/$c$.
Without this capability, the flavour physics programme would be limited by kaon-pion
misidentification in multi-body $B$ decays.
Candidate technologies include LGAD-based TOF layers or TORCH-like detectors.
The calorimetry, discussed in Section~\ref{sec:calo}, also uses timing in the
dual-readout mode for longitudinal shower profiling.

\subsection{Muon Collider}
\label{sec:mucoll}

A Muon Collider at centre-of-mass energies of several TeV~\cite{MuColl_CDR} faces a
unique background environment: muons decay continuously along the accelerating and
storage ring, producing a flux of electrons, photons, and neutrons (Beam-Induced
Background, BIB) that arrives at the detector \textit{asynchronously} with the
colliding bunch.
The BIB rate is several orders of magnitude higher than the physics rate, and its
particles have a broad time distribution spanning $\pm$10\,ns around the collision.

Timing is the primary tool for BIB suppression.
A gate of $\pm$150\,ps around the collision time, feasible with 20--30\,ps per-hit
resolution, removes $>$99\% of BIB hits from the tracker.
Without this, the detector occupancy would be nearly 100\% and event reconstruction
would be impossible.
The vertex detector targets 20--30\,ps, while the calorimeter aims for 80\,ps per cell
(Section~\ref{sec:calo}).

\subsection{FCC-hh}
\label{sec:fcchh}

The hadron mode of the FCC (FCC-hh)~\cite{FCChh_CDR} targets a centre-of-mass energy of 100\,TeV with
a pile-up of $\sim$1000 simultaneous proton-proton interactions per bunch crossing.
At such pile-up densities, spatial tracking alone becomes increasingly challenged in resolving primary vertices along the beam axis. Design studies indicate that per-hit time resolutions approaching or below 10 ps would substantially improve vertex separation in a combined “Time-Z” reconstruction, where both spatial and temporal coordinates are used for track association. 
Reaching this level of performance motivates continued development of advanced CMOS nodes (28 nm and beyond) together with sensor concepts targeting single-hit resolutions well below 10 ps. Values in the few-picosecond regime are currently considered long-term R\&D objectives rather than demonstrated detector-level performance under realistic radiation and power constraints.

The timing requirements of these facilities are compared in Table~\ref{tab:future}.

\begin{table}[t!]
\vspace{.2cm}
\begin{center}
\begin{tabular}{l l l l}
\toprule
Facility & Pile-up / BIB challenge & $\sigma_t$ target & Primary motivation \\
\midrule
FCC-ee      & Low (clean $e^+e^-$)    & $\sim$10\,ps  & PID ($K/\pi$), dual-readout cal. \\
Muon Coll. & BIB ($>$99\% rejection) & 20--30\,ps    & Tracker: BIB gating \\
Muon Coll. & BIB                     & $\sim$80\,ps  & Calorimeter: BIB gating \\
FCC-hh      & $\mu\approx1000$        & $<$10\,ps     & Vertex reconstruction \\
\bottomrule
\end{tabular}
\caption{Timing requirements for far-future facilities.
\label{tab:future}}
\end{center}
\vspace{-.3cm}
\end{table}

\section{Timing in future Calorimeters}
\label{sec:calo}

Calorimeters have traditionally sacrificed time resolution for energy resolution;
modern systems are reversing this priority, treating timing as an essential tool for
shower cleaning, pile-up rejection, and longitudinal profiling.

\subsection{LHCb PicoCal}
\label{sec:picocal}

The LHCb calorimeter Upgrade~II (PicoCal) is a multi-technology proposal targeting
$\sim$20\,ps per cluster.
It combines two readout geometries:
\begin{itemize}
  \item \textit{SpaCal} (Spaghetti Calorimeter): tungsten/lead absorber with
  scintillating fibres read out by SiPMs.
  \item \textit{Shashlik}: scintillator/lead stacks with wavelength-shifting fibres,
  also SiPM-read.
\end{itemize}
The goal is to separate overlapping electromagnetic showers in the high-occupancy
central region of LHCb at Upgrade~II luminosities.

\subsection{Crilin for the Muon Collider}
\label{sec:crilin}

The Crilin crystal calorimeter~\cite{Crilin} has been developed specifically for the
BIB environment of the Muon Collider.
It uses PbF$_2$ (pure Cherenkov) or ultra-fast PbWO$_4$ crystals read out by SiPMs.
Test beam measurements have demonstrated $<$25\,ps for electrons above 3\,GeV.
Since BIB particles arrive with a broad time distribution, while physics products
arrive within $\pm$150\,ps of the collision, a per-cell resolution of $\sim$25--50\,ps
provides effective BIB gating in the calorimeter.
This is the calorimetric analogue of the tracking BIB gate described in
Section~\ref{sec:mucoll}.

\subsection{Dual-readout calorimetry: IDEA / FCC-ee}
\label{sec:dualread}

The dual-readout approach (IDEA detector at FCC-ee, HiDRa and ASPIDES
projects~\cite{IDEA_dualread}) reads out both Scintillation and Cherenkov light from
the same fibre matrix simultaneously.
The Scintillation signal integrates over the full hadronic shower; the Cherenkov signal
responds only to the electromagnetic component.
The ratio of the two signals, combined with a model of hadronic shower development,
determines the electromagnetic fraction and hence corrects for the intrinsic
non-compensation of the sampling calorimeter.

Timing adds a third dimension: the Cherenkov signal is prompt ($<$1\,ns), while
scintillation is delayed by the fluorescence lifetime ($\sim$2--5\,ns for fast
crystals, longer for doped fibres).
By measuring the arrival time of each fibre signal with $<$100\,ps resolution,
the system reconstructs the longitudinal position of the shower centre without
physical longitudinal segmentation.
The readout electronics target $<$100\,ps per fibre using SiPMs coupled to ASICs such
as Citiroc or dedicated TDC chips.

Table~\ref{tab:calo} summarises the calorimeter timing systems described in this section.

\begin{table}[t!]
\vspace{.2cm}
\begin{center}
\begin{tabular}{l l l l}
\toprule
System & Technology & $\sigma_t$ goal & Primary timing role \\
\midrule
LHCb PicoCal & SpaCal / Shashlik + SiPM & $\sim$20\,ps per cluster & Shower separation \\
Crilin (Mu Coll.) & PbF$_2$/PbWO$_4$ + SiPM & $<$25\,ps for $e^->$3\,GeV & BIB gating \\
IDEA (FCC-ee) & Dual-readout fibres + SiPM & $<$100\,ps per fibre & Longitudinal profiling \\
Mu Coll.\ ECal & Fast crystals + SiPM & $\sim$80\,ps per cell  & BIB gating \\
\bottomrule
\end{tabular}
\caption{Calorimeter timing systems: present construction and future plans.
\label{tab:calo}}
\end{center}
\vspace{-.3cm}
\end{table}

\section{Timing in future Gaseous and Cherenkov detectors}
\label{sec:gas}

Gaseous and Cherenkov timing technologies retain important roles in the future landscape,
particularly where large areas must be covered at low cost, or where particle
identification is required over macroscopic flight paths.

\subsection{TORCH at LHCb Upgrade II}
\label{sec:torch}

TORCH~\cite{TORCH} will be installed in LHCb for Upgrade~II.
It uses 1\,cm thick fused-silica quartz plates as Cherenkov radiators and
Micro-Channel Plate PMTs (MCP-PMT) for photon detection.
The system reconstructs the time-of-flight of a charged particle over a flight path
of 9.5\,m downstream of the interaction point.
A per-photon resolution of $\sim$70\,ps from the MCP-PMTs combines over $\sim$30
detected photons per track to yield a per-track resolution of $\sim$15\,ps.
This provides $3\sigma$ $K/\pi$ separation up to 10\,GeV/$c$ and $p/K$ separation up
to 15\,GeV/$c$, complementing the silicon TOF layers for high-momentum PID.
Readout is provided by waveform-digitizer ASICs of the ASOC/SAMPIC family.

\subsection{PICOSEC Micromegas}
\label{sec:picosec}

The PICOSEC Micromegas~\cite{PICOSEC} is a hybrid gaseous detector combining a
Cherenkov radiator (MgF$_2$) and a photocathode (CsI) with a two-stage Micromegas
amplification gap.
A charged particle generates Cherenkov photons in the MgF$_2$ window; these liberate
photoelectrons from the CsI photocathode, which are pre-amplified in the first
(``drift'') gap and further amplified in the Micromegas gap.
The resulting signal has a rise time of $\sim$1\,ns, yielding a consistently measured
resolution of $<$25\,ps for MIPs.

PICOSEC is proposed for large-area forward timing walls where the cost of silicon
would be prohibitive.
Its radiation hardness exceeds that of conventional MRPCs, and the gas amplification
gain can be tuned to compensate for photocathode ageing.

\subsection{4D-RICH}
\label{sec:4drich}

Several projects are exploring the addition of timing to Ring-Imaging Cherenkov (RICH)
detectors.
By coupling an MCP-PMT to a Timepix4 pixelated anode~\cite{Bolzonella2026}, single-photon timing of
$\sim$30\,ps becomes achievable, enabling the per-photon timestamps to be used as
an additional filter against background photons.
In environments such as ALICE~3 or LHCb Upgrade~II, where the occupancy per ring
is high, rejecting photons whose arrival time is inconsistent with the track time
reduces the combinatorial background in ring finding significantly.
The concept is being developed under the name ``4D-RICH'' project.

Table~\ref{tab:gas} summarises gaseous and Cherenkov timing systems past and future.

\begin{table}[t!]
\vspace{.2cm}
\begin{center}
\begin{tabular}{l l l l l}
\toprule
System & Technology & $\sigma_t$ & Application & Status \\
\midrule
ALICE TOF (Run 1--2) & MRPC + NINO + HPTDC & $\sim$80\,ps & Hadron PID & Operational \\
Belle~II TOP         & Quartz + MCP-PMT    & $<$50\,ps/track & $K/\pi$ PID & Operational \\
TORCH (LHCb Upgrade II) & Quartz + MCP-PMT & $\sim$15\,ps/track & $K/\pi$, $p/K$ & R\&D / construction \\
PICOSEC Micromegas   & MgF$_2$ + CsI + gas & $<$25\,ps & Large-area TOF wall & R\&D \\
4D-RICH              & MCP-PMT + Timepix4  & $\sim$30\,ps/photon & RICH enhancement & R\&D \\
\bottomrule
\end{tabular}
\caption{Gaseous and Cherenkov timing systems.
\label{tab:gas}}
\end{center}
\vspace{-.3cm}
\end{table}

\section{Timing in future Satellite Experiments}
\label{sec:space}

Space-borne cosmic-ray detectors impose constraints qualitatively different from
those of ground-based particle physics experiments, and timing in space has its own
unique requirements.
The primary drivers are: particle identification and isotope separation via TOF,
charge measurement via $dE/dx$, directionality determination, and a strict power
budget typically below 20\,mW/cm$^2$.

Experiments such as AMS-02 (Section~\ref{sec:pmt_physics}), DAMPE, and future planned missions use
TOF measurements for:
\begin{itemize}
\item PID and isotope separation: distinguishing, for example, $^3$He from
$^4$He on the basis of their velocity at the same rigidity.
The mass-squared difference scales as $(M_A^2-M_B^2)/p^2$; for helium isotopes at
a few GV of rigidity, this requires a TOF resolution of $\mathcal{O}(100)$\,ps over
a flight path of $\sim$1\,m.
\item Directionality and self-veto: as in AMS-02, the top-to-bottom time
difference between two silicon tracker layers determines whether a particle is
entering from above (cosmic ray) or below (secondary), allowing a single silicon layer
to function as a directional detector without a dedicated TOF scintillator.
This is a direct application of 4D silicon sensing to the space environment.
\end{itemize}

Future satellite missions are evaluating thin LGAD layers as a replacement for
scintillator/PMT TOF paddles, offering lower mass, lower voltage operation, and
the possibility of fine-grained spatial information at the same time.
The power constraint of $<$20\,mW/cm$^2$ is compatible with the low-power designs
being developed for ALICE~3, making these two communities natural collaborators.

\section{Timing ASICs}
\label{sec:asics}

The development of the timing ASIC is arguably the most technically challenging
aspect of the transition to silicon timing. While some design blocks are common for SiPMs and LGADs, 
ASICs for SiPM and LGAD readout differ substantially  in several aspects such as the analogue front-end design
(SiPM signals, $\sim$1--10\,pF capacitance, $\sim$1--4\,mA peak current, are
qualitatively different from LGAD signals, $\sim$0.1--1\,pF, $\sim$10--100\,$\mu$A
peak current) and specialized circuits such as the SiPM dark count suppression.

Table~\ref{tab:asics} summarises the principal chips discussed below.

\subsection{The NA62 GigaTracker}
\label{sec:asics_early}
\begin{figure}[!ht]
  \vspace{.2cm}
  \centering
  \includegraphics[width=.85\linewidth]{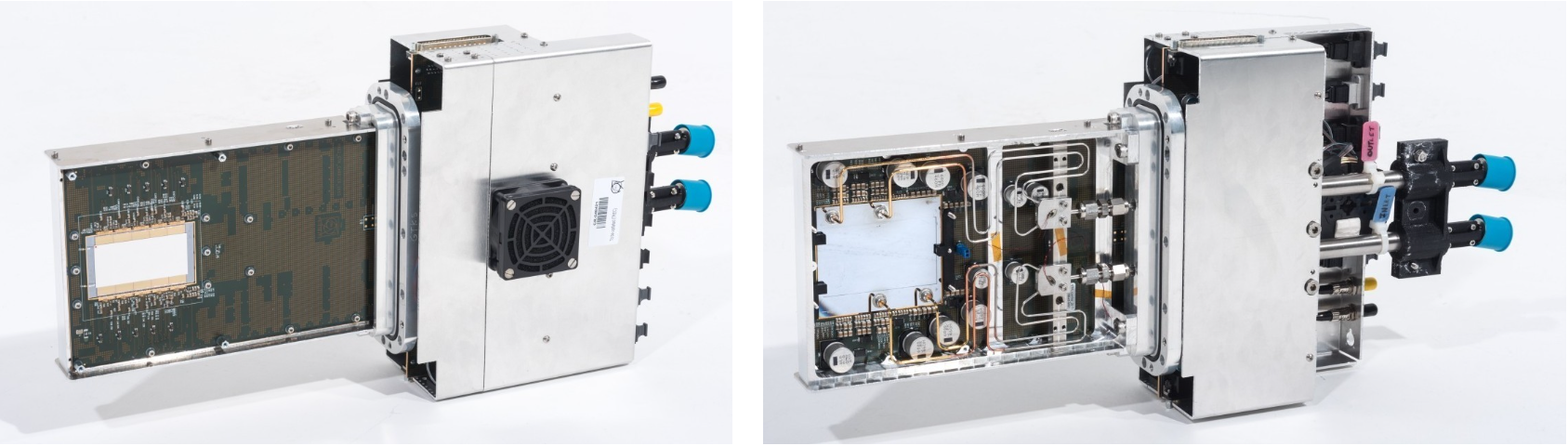}
  \caption{Photographs of a NA62 GigaTracker station, showing the sensor side (left)
  and the cooling plate side (right). The station houses a $60.8\times27$\,mm$^2$
  hybrid pixel detector with $300\times300\,\mu$m$^2$ pixels, read out by the
  TDCPix ASIC fabricated in 130\,nm CMOS. Operating in the full NA62 kaon beam at
  $\sim$1\,MHz/mm$^2$, the GigaTracker was the first demonstration that sub-100\,ps
  timing could be integrated directly into a high-rate pixel detector, establishing
  the blueprint for all subsequent timing ASICs~\cite{Neri2012}.
  \label{fig:gigatracker}}
\end{figure}

The TDCPix ASIC developed for the NA62 GigaTracker~\cite{Neri2012} (Fig.~\ref{fig:gigatracker}) is widely regarded
as the first of modern silicon timing ASICs.
Fabricated in 130\,nm CMOS, it provides Time-of-Arrival (ToA) measurements with
$\sim$100\,ps time resolution across a $60\times27$\,mm$^2$ die containing $18\,000$ pixels
of $300\times300\,\mu$m$^2$.
The GigaTracker operated at a beam rate of $\sim$1\,MHz/mm$^2$ in the NA62 kaon beam,
demonstrating for the first time that sub-100\,ps timing~\cite{Aglieri2019} could be integrated directly
into a high-rate pixel detector at moderate power consumption.

\subsection{ASICs for the ATLAS and CMS timing layers at the HL-LHC}
\label{sec:asics_hllhc}

\paragraph{ETROC (CMS ETL)}
ETROC~\cite{ETROC} is fabricated in 65\,nm CMOS and provides Time-of-Arrival (ToA)
and Time-over-Threshold (ToT) measurements for the CMS LGAD timing layer.
The ToT information enables time-walk correction on-chip.
Electronics jitter is targeted at $\sim$25\,ps.
The ETROC2 prototype has demonstrated full functionality with LGAD sensors in test
beams.

\paragraph{ALTIROC (ATLAS HGTD)}
ALTIROC~\cite{ALTIROC} is fabricated in 130\,nm CMOS and reads out the ATLAS LGAD disks.
It uses a similar ToA + ToT architecture and targets a comparable electronics
contribution to the overall 25--35\,ps system resolution.

\paragraph{TOFHIR2 (CMS BTL)}
The TOFHIR2 ASIC~\cite{TOFHIR2} is designed for the CMS LYSO+SiPM barrel layer.
It operates in 130\,nm CMOS and reads out both timing and charge.
A dedicated dark-count cancellation scheme suppresses SiPM thermal noise hits.
The electronics contribution targets 30\,ps at beginning of life, rising to $\sim$60\,ps
as radiation damage increases SiPM dark count rates at end of life.

\subsection{ASICs for EIC}
\label{sec:asics_rsd}

AC-LGAD (RSD) sensors require ASICs that measure both ToA and pulse amplitude at each
electrode, because the hit position is reconstructed from the charge ratio between
neighbouring electrodes.
This imposes a different analogue architecture from the standard LGAD readout.

\paragraph{EICROC}
The EICROC~\cite{EICROC} is a 130\,nm ASIC developed by a consortium including
BNL and Fermilab for the EIC forward TOF.
It measures ToA and charge simultaneously and is designed for $0.5\times0.5$\,mm$^2$
pixels at the high hit-rate of the EIC forward region.

\paragraph{FCFD}
The FCFD~\cite{FCFD} (Fermilab, 65\,nm CMOS) uses a constant-fraction discriminator
to measure signal arrival time, avoiding the need for time-walk correction by design.
It achieves $\sim$10\,ps electronics contribution for LGAD signals above 20\,fC,
aimed at barrel TOF applications.

\subsection{Next-generation frontier: 28\,nm ASICs}
\label{sec:asics_28nm}

\paragraph{TIMESPOT1}
TIMESPOT1~\cite{TIMESPOT1}, developed by INFN, was the first 28\,nm demonstrator ASIC
for in-pixel TDC integration.
It achieves a time resolution of $\sim$7\,ps per pixel for $55\times55\,\mu$m$^2$
pixels -- demonstrating that the intrinsic limit of a thin 3D silicon sensor can be
reached at the ASIC level.
The primary limitation is power density of $\sim$1.5--2\,W/cm$^2$, which exceeds
the sustainable limit for large-area detectors and motivates the resource-sharing
strategies of successor chips.

\paragraph{PICOPIX}
PICOPIX~\cite{PICOPIX} is developed within the
Timepix family at CERN.
It implements the \textit{Analogue Island} concept, sharing TDC resources among groups
of pixels and using power-gating techniques to reduce the average power to a
sustainable level while maintaining $\sim$20\,ps timing.
PICOPIX is the baseline readout ASIC for the LHCb VELO Upgrade~II.

\subsection{Alternative paths: waveform digitizers}
\label{sec:asics_waveform}

For detectors with fewer channels (MCP-PMT arrays, small SiPM matrices), where
the data volume is manageable, waveform digitizers offer the best possible timing.
The ASOC and SAMPIC~\cite{SAMPIC} chips sample the full pulse shape at
$\sim$1--8\,GS/s (equivalent to a GHz oscilloscope on a chip), allowing
offline analysis of the full pulse shape for sub-10\,ps timing.
The cost is high power ($\sim$1\,W per channel) and large data volume, restricting
these chips to low-channel-count applications such as TORCH (Section~\ref{sec:torch}).
EICROC offers also this capability, sampling the waveform in several points. 

\begin{table}[t!]
\vspace{.2cm}
\begin{center}
\begin{tabular}{l l l l l}
\toprule
ASIC & Process & Electronics jitter & Sensor type & Application \\
\midrule
ETROC       & 65\,nm     & $\sim$25\,ps    & LGAD   & CMS MTD ETL \\
ALTIROC     & 130\,nm    & $\sim$30\,ps    & LGAD   & ATLAS HGTD \\
TOFHIR2     & 130\,nm    & 30--60\,ps      & SiPM + Crystal & CMS MTD BTL \\
EICROC      & 130\,nm    & $\sim$30\,ps    & AC-LGAD & EIC ePIC TOF \\
FCFD        & 65\,nm     & $\sim$10\,ps    & LGAD / AC-LGAD & EIC ePIC TOF \\
TIMESPOT1   & 28\,nm     & $\sim$7\,ps     & 3D Trench / LGAD & Demonstrator \\
PICOPIX     & 28\,nm     & $\sim$20\,ps    & 3D Trench / LGAD & LHCb VELO II \\
SAMPIC/ASOC & 130--65\,nm & $<$10\,ps      & MCP-PMT / SiPM  & TORCH, small systems \\
\bottomrule
\end{tabular}
\caption{Summary of principal timing ASICs.
\label{tab:asics}}
\end{center}
\vspace{-.3cm}
\end{table}

\section{Future Challenges}
\label{sec:challenges}

The path from today's 25--50\,ps systems to the $<$10\,ps required by FCC-hh and
FCC-ee is not merely evolutionary; it requires solutions to several fundamental
problems.

\paragraph{Low-power timing ASICs}
Current 28\,nm demonstrators consume $\sim$1.5--2\,W/cm$^2$, a factor of five above
what can be removed by evaporative CO$_2$ cooling without unacceptable material budget.
Achieving $<$0.5\,W/cm$^2$ at $\sim$10\,ps requires either a radical reduction i power through novel circuit topologies  or architectural
innovations such as deep resource sharing.

\paragraph{Radiation hardness to $10^{16}$\,n$_\text{eq}$/cm$^2$}
Standard LGADs lose their gain layer at fluences above $\sim$2$\times10^{15}$\,n$_\text{eq}$/cm$^2$.
The \textit{Compensated LGAD} concept~\cite{Compensated_LGAD} co-implants donor atoms
to partially compensate the acceptor removal, extending the useful gain to fluences
approaching $10^{16}$\,n$_\text{eq}$/cm$^2$.
Alternatively, 3D trench sensors (which have no gain layer to destroy) can operate
at the highest fluences with $\sim$10\,ps intrinsic timing, but require more complex
fabrication and the temporal resolution degrades rapidly as the pixel size increases above the presently used 50-55\,$\mu$m.

\paragraph{Picosecond clock distribution}
A timing resolution of 10\,ps  requires distributing a clock with $<$5\,ps jitter across millions of channels.
Optical clock distribution schemes, which avoid the impedance mismatches of electrical
networks, are under investigation.

\paragraph{Thermal management}
Even at 0.5\,W/cm$^2$, a large tracking system of $\mathcal{O}(10)$\,m$^2$ dissipates
$\mathcal{O}(50)$\,kW.
Evaporative CO$_2$ cooling is the current standard, but its mass ($\sim$1\% $X_0$ per
cooling circuit) competes with the material budget, particularly in inner tracker layers.

\section{Conclusions}
\label{sec:conclusions}

The history of timing in particle physics is a story of increasing integration:
from a PMT behind a scintillator bar, to a dedicated timing layer of silicon pixels,
to time as a fourth coordinate measured at every hit in a tracking detector.

Table~\ref{tab:evolution} summarises the four generations in terms of their key
performance figures.

\begin{table}[t!]
\vspace{.2cm}
\begin{center}
\begin{tabular}{l l l l l}
\toprule
Era & Technology & Resolution & Channels & Integration \\
\midrule
1990--2005 & Scint.\ + PMT + NIM crates & 100--200\,ps & $10^2$--$10^5$ & Dedicated subsystem \\
2005--2025 & SiPM / LGAD + ASIC        & 25--80\,ps   & $10^5$--$10^7$ & Timing layer (3+1) \\
2025--2040 & LGAD / 3D + 28\,nm ASIC   & 10--30\,ps   & $10^6$--$10^8$ & 4D at every hit \\
2040+      & Monolithic LGAD  & $<$10\,ps    & $10^8$--$10^9$ & Ubiquitous 4D \\
\bottomrule
\end{tabular}
\caption{The four generations of timing in particle physics.
\label{tab:evolution}}
\end{center}
\vspace{-.3cm}
\end{table}

The silicon revolution, enabled by SiPM, LGAD, and timing ASICs, transformed
timing from a specialised technique into a mainstream detector capability.
The systems under construction at the HL-LHC -- CMS MTD, ATLAS HGTD, and CMS HGCAL --
demonstrate the maturity of this technology at the million-channel scale with 25--50\,ps
resolution.
The EIC, LHCb VELO Upgrade~II, and ALICE~3 push into new regimes, with LHCb providing
the first example of full 4D tracking in a collider environment.
Future facilities, from the FCC-ee precision factory to the extreme environment of the
Muon Collider and the enormous pile-up of the FCC-hh, require a further leap to
$\sim$10\,ps, setting a clear and demanding R\&D agenda.
The dominant challenges are power, radiation hardness, and clock distribution -- all
solvable in principle, but requiring sustained investment in sensor and ASIC development
over the coming decade. The transition from today’s 25--50 ps systems to the $<$10 ps regime is constrained as much by fundamental fluctuation limits as by engineering ingenuity.

\section*{Acknowledgements}
The author thanks the many colleagues in the LGAD, SiPM, and timing ASIC communities
whose work is reviewed here. This paper was presented at TIPP2026, Mumbai.

{}


\begin{thebibliography}{}

\bibitem{Acosta2005}
D.~Acosta et al.\ (CDF Collaboration),
\textit{The CDF time-of-flight detector},
Nucl.\ Instrum.\ Meth.\ A \textbf{518} (2004) 605.

\bibitem{Fanti1999}
V.~Fanti et al.\ (NA48 Collaboration),
\textit{A new measurement of direct CP violation in two pion decays of the neutral kaon},
Phys.\ Lett.\ B \textbf{465} (1999) 335.

\bibitem{Alavi-Harati1999}
A.~Alavi-Harati et al.\ (KTeV Collaboration),
\textit{Observation of direct CP violation in $K_{S,L}\to\pi\pi$ decays},
Phys.\ Rev.\ Lett.\ \textbf{83} (1999) 22.

\bibitem{Aguilar2002}
M.~Aguilar et al.\ (AMS Collaboration),
\textit{The Alpha Magnetic Spectrometer (AMS) on the International Space Station},
Phys.\ Rep.\ \textbf{366} (2002) 331.

\bibitem{Fukuda1998}
Y.~Fukuda et al.\ (Super-Kamiokande Collaboration),
\textit{Evidence for an oscillatory signature in atmospheric neutrino oscillation},
Phys.\ Rev.\ Lett.\ \textbf{81} (1998) 1562.

\bibitem{Anghinolfi2004}
F.~Anghinolfi et al.,
\textit{NINO: an ultra-fast and low-power front-end amplifier-discriminator ASIC
designed for the multigap resistive plate chamber},
Nucl.\ Instrum.\ Meth.\ A \textbf{533} (2004) 183.

\bibitem{Christiansen2004}
J.~Christiansen,
\textit{HPTDC High Performance Time to Digital Converter},
CERN/EP-MIC, v2.2 (2004).

\bibitem{ALICE_TOF}
ALICE Collaboration,
\textit{ALICE Time-Of-Flight system (TOF): Technical Design Report},
CERN/LHCC 2000-012 (2000).

\bibitem{Aubert2002}
B.~Aubert et al.\ (BaBar Collaboration),
\textit{The BaBar detector},
Nucl.\ Instrum.\ Meth.\ A \textbf{479} (2002) 1.

\bibitem{BelleII_TDR}
Belle~II Collaboration,
\textit{The Belle II Physics Book},
PTEP \textbf{2019} (2019) 123C01.

\bibitem{Renker2009}
D.~Renker and E.~Lorenz,
\textit{Advances in solid state photon detectors},
JINST \textbf{4} (2009) P04004.

\bibitem{Pellegrini2014}
G.~Pellegrini et al.,
\textit{Technology developments and first measurements of Low Gain Avalanche Detectors
(LGAD) for high energy physics applications},
Nucl.\ Instrum.\ Meth.\ A \textbf{765} (2014) 12.

\bibitem{Cartiglia2015}
N.~Cartiglia et al.,
\textit{Design optimization of Ultra-Fast Silicon Detectors},
Nucl.\ Instrum.\ Meth.\ A \textbf{796} (2015) 141.

\bibitem{Sadrozinski2018}
H.~Sadrozinski, A.~Seiden, N.~Cartiglia,
\textit{4D tracking with ultra-fast silicon detectors},
Rep.\ Prog.\ Phys.\ \textbf{81} (2018) 026101.

\bibitem{Moll2018}
M.~Moll,
\textit{Acceptor removal -- Displacement damage effects involving the shallow
acceptor doping of p-type silicon devices},
Nucl.\ Instrum.\ Meth.\ A \textbf{924} (2019) 97.

\bibitem{Mandurrino2019}
M.~Mandurrino et al.,
\textit{Demonstration of 200-, 100-, and 50-$\mu$m pitch resistive AC-coupled
silicon detectors (RSD) with 100\% fill factor for 4D particle tracking},
IEEE Electron Device Lett.\ \textbf{40} (2019) 1780.

\bibitem{Aglieri2019}
G.~Aglieri Rinella et al.,
\textit{The GigaTracker: A sub-100 ps time-stamping silicon pixel detector
for the NA62 experiment at CERN},
JINST \textbf{14} (2019) P07010.

\bibitem{Neri2012}
I.~Neri et al.,
\textit{The TDCPix ASIC for the NA62 GigaTracker},
JINST \textbf{7} (2012) C01001.

\bibitem{CMS_MTD_TDR}
CMS Collaboration,
\textit{A MIP timing detector for the CMS Phase-2 upgrade},
CERN/LHCC 2019-003, CMS-TDR-020 (2019).

\bibitem{TOFHIR2}
G.~Cerminara et al.,
\textit{The TOFHIR2 ASIC for the CMS Barrel Timing Layer},
JINST \textbf{17} (2022) C03018.

\bibitem{ETROC}
CMS Collaboration,
\textit{ETROC: the readout ASIC for the CMS ETL},
CMS-DN-21-001 (2021).

\bibitem{ATLAS_HGTD_TDR}
ATLAS Collaboration,
\textit{Technical Design Report: A High-Granularity Timing Detector for the ATLAS Phase-II Upgrade},
CERN/LHCC 2020-007, ATL-TDR-31 (2020).

\bibitem{FCFD}
S.~Xie et al.,
\textit{Design and performance of the Fermilab Constant Fraction Discriminator ASIC},
Nucl.\ Instrum.\ Meth.\ A \textbf{1056} (2023) 168655.

\bibitem{ALTIROC}
C.~Agapopoulou et al.,
\textit{Performance of a front-end prototype ASIC for the ATLAS High Granularity timing detector},
JINST \textbf{18} (2023) P08019.

\bibitem{CMS_HGCAL_TDR}
CMS Collaboration,
\textit{The Phase-2 Upgrade of the CMS Endcap Calorimeter},
CERN/LHCC 2017-023, CMS-TDR-019 (2017).

\bibitem{PIONEER_CDR}
PIONEER Collaboration,
\textit{PIONEER: Studies of rare pion decays},
arXiv:2203.01981 (2022).

\bibitem{EIC_CDR}
A.~Accardi et al.,
\textit{Electron-Ion Collider: The Next QCD Frontier},
Eur.\ Phys.\ J.\ A \textbf{52} (2016) 268.

\bibitem{EICROC}
EIC Collaboration,
\textit{EICROC: the readout ASIC for EIC AC-LGAD TOF},
ePIC technical note (2023).

\bibitem{LHCb_VELO_II}
LHCb Collaboration,
\textit{LHCb VELO Upgrade II: Expression of Interest},
CERN-LHCC-2022-002 (2022).

\bibitem{PICOPIX}
O.A.~de~Aguiar~Francisco,
\textit{The LHCb VELO Upgrade~II: Design and development of the readout electronics},
Nucl.\ Instrum.\ Meth.\ A \textbf{1063} (2024) 169238.

\bibitem{ALICE3_LoI}
ALICE Collaboration,
\textit{Letter of intent for ALICE 3: A next-generation heavy-ion experiment at the LHC},
arXiv:2211.02491 (2022).

\bibitem{FCCee_CDR}
FCC Collaboration,
\textit{FCC-ee: The Lepton Collider},
Eur.\ Phys.\ J.\ Spec.\ Top.\ \textbf{228} (2019) 261.

\bibitem{FCChh_CDR}
FCC Collaboration,
\textit{FCC-hh: The Hadron Collider},
Eur.\ Phys.\ J.\ Spec.\ Top.\ \textbf{228} (2019) 755.

\bibitem{MuColl_CDR}
C.~Aime et al.,
\textit{Muon Collider Physics Summary},
arXiv:2203.07256 (2022).

\bibitem{TIMESPOT1}
M.~Milanesio et al.,
\textit{TIMESPOT1: A 28-nm CMOS Pixel ASIC for 4D tracking in high-hit rate environments},
JINST \textbf{17} (2022) C11004.

\bibitem{SAMPIC}
D.~Breton et al.,
\textit{The SAMPIC waveform and time-to-digital converter},
IEEE NSS/MIC (2014).

\bibitem{Crilin}
E.~Diociaiuti et al.,
\textit{Crilin: A CRystal calorImeter with Longitudinal INformation for future
lepton colliders},
arXiv:2204.12232 (2022).

\bibitem{IDEA_dualread}
A.~Lucchini et al.,
\textit{New perspectives on segmented crystal calorimeters for future colliders},
JINST \textbf{15} (2020) P11005.

\bibitem{TORCH}
T.~Gys et al.,
\textit{The TORCH detector: R\&D towards a large-scale device},
Nucl.\ Instrum.\ Meth.\ A \textbf{952} (2020) 161783.

\bibitem{PICOSEC}
J.~Bortfeldt et al.,
\textit{PICOSEC: Charged particle timing at sub-25 picosecond precision with a
Cherenkov based Micromegas detector},
Nucl.\ Instrum.\ Meth.\ A \textbf{903} (2018) 317.

\bibitem{Bolzonella2026}
R.~Bolzonella et al.,
\textit{Development and characterization of hybrid MCP-PMT with embedded Timepix4 ASIC used as pixelated anode},
Nucl.\ Instrum.\ Meth.\ A \textbf{1082} (2026) 170965.

\bibitem{FAST3}
M.~Ferrero et al.,
\textit{FAST3 ASIC: an analog front-end with 30\,ps resolution, designed to readout thin Low Gain Avalanche Diodes},
JINST \textbf{20} (2025) C03007.

\bibitem{Compensated_LGAD}
V.~Sola et al.,
\textit{A compensated design of the LGAD gain layer},
Nucl.\ Instrum.\ Meth.\ A \textbf{1041} (2022) 167228.

\end{thebibliography}
\end{document}